%% file: main.tex
\documentclass[twocolumn]{aastex63}
\usepackage{graphicx}

\usepackage{hyperref}
\usepackage{amssymb}
\usepackage{amsmath}
\usepackage{url}

\usepackage{color, colortbl}
\usepackage[version=4]{mhchem}

\usepackage{xcolor}

\usepackage{color, colortbl}
\definecolor{Gray}{gray}{0.75}
\usepackage{natbib}

\begin{document}
\title{The Detection-vs-Retrieval Challenge: Titan as an Exoplanet}

\author[0000-0002-8052-3893]{Prajwal Niraula}
\affiliation{Department of Earth, Atmospheric and Planetary Sciences, MIT, 77 Massachusetts Avenue, Cambridge, MA 02139, USA}
\affiliation{Department of Physics and Kavli Institute for Astrophysics and Space Research, Massachusetts Institute of Technology, Cambridge, MA
02139, USA}
\email{Corresponding author Email: pniraula@mit.edu}

\author[0000-0003-2415-2191]{Julien de Wit}
\affiliation{Department of Earth, Atmospheric and Planetary Sciences, MIT, 77 Massachusetts Avenue, Cambridge, MA 02139, USA}

\author[0000-0002-7691-6926]{Robert J. Hargreaves} 
\affil{Center for Astrophysics \textbar Harvard \& Smithsonian, Atomic and Molecular Physics Division, Cambridge, MA 02138, USA}

\author[0000-0003-4763-2841]{Iouli E. Gordon} 
\affil{Center for Astrophysics \textbar Harvard \& Smithsonian, Atomic and Molecular Physics Division, Cambridge, MA 02138, USA}

\author[0000-0002-7853-6871]{Clara Sousa-Silva}
\affil{Physics Program, Bard College, 30 Campus Road, Annandale-On-Hudson, New York 12504, USA}


\begin{abstract}
Cassini observations of Titan's atmosphere are exemplary benchmarks for exoplanet atmospheric studies owing to their precision, spectral coverage, and our independent knowledge of Titan. We carry out atmospheric retrievals of Titan’s transmission spectrum to test the sensitivity of interpretations to the range of molecular species included in the retrieval (i.e., set \textit{a priori} as ``detectable''). We find that hydrocarbons (HCs) such as ethane or iso-propane can be ``retrieved'' from Titan's 3.3-$\mu$m band. However, given that a multitude of HC species exhibit overlapping features in this region, such retrievals cannot be claimed as bonafide ``detections", particularly given the limited accuracy and availability of associated cross-sections. The choices of HCs used in the retrieval process can vary the retrieved abundance of key absorbers like methane by $\sim$0.5 dex (a factor of $\sim$3). This underscores the broader issue that beyond the possible misidentification of molecular features (e.g., the inventory debate surrounding dimethyl sulfide, DMS, in K2-18 b), the inventory of molecular species, often dictated by computational considerations, can bias the retrieved atmospheric parameters. We thus recommend a sensitivity analysis to assess the dependencies of atmospheric inferences on such \textit{a priori} selections, in tandem with complementary information (e.g., chemistry models). Finally, we show an independent path to constrain the dominant atmospheric constituent, even when lacking observable absorption features (e.g., H$_2$ and N$_2$) through scale height.
\end{abstract}


\section{Introduction} 
\label{sec:intro}
Titan hosts a complex and high metallicity atmosphere, blanketed by haze \citep{kuiper1944}. The pervasive haze gives Titan its characteristic orange hue \citep{horst2017}. The Visual and Infrared Mapping Spectrometer \citep[VIMS,][]{Brown2004} onboard Cassini performed high-fidelity observations of Titan's atmosphere using solar occultations \citep{robinson2014}. Cassini's observations represent an aspirational precision goal for future exoplanet atmospheric studies. They provide spectral coverage and resolution comparable to that of JWST/PRISM, but are roughly an order of magnitude more precise. The data hence presents an opportunity to investigate arising roadblocks in interpreting exoplanet data from JWST and beyond, as extracting robust scientific inferences is increasingly limited by our ability to build high-fidelity models. Building a high fidelity model requires careful consideration from cloud parameterization to their properties \citep{line2016, welbanks2019} including patchiness and differences across terminators \citep{welbanks2022, espinoza2024}, understanding the hydrogen to helium ratio \citep{deWit2025Helium}, stellar contamination \citep{rackham2018, rackham2023} to day/night side temperature asymmetries \citep{caldas2019, macDonald2021}, among others.

In this study, we leverage Titan's precise transmission spectrum and our existing knowledge of its atmosphere to investigate the strengths/limitations of exoplanet atmospheric retrievals, with a particular focus on hydrocarbons (HCs). As a well-studied system, Titan serves as a test bed for exoplanetary atmospheric retrieval, illustrating how overlapping HC features challenge standard detection frameworks \citep{madhusudhan2009, benneke2013, dewit2013}.  This focus is timely owing to existing concerns associated with the possible misinterpretation of molecular features such as with DMS in K2-18\,b \citep{Madhusudhan2025}, a sensitive result given its astrobiological implications. Subsequent studies have raised concerns; \citep{Thorngren2025} pointed to  an optimistic interpretation of the Bayes factor that would otherwise mean a marginal detection, \citet{Luque2025} and \citet{Schmidt2025} found that the evidence for DMS weakens, rather than strengthens, when data covering a broader spectral range are considered, and \citet{Welbanks2025} highlighted common pitfalls in retrieval frameworks, demonstrating that alternative molecules, such as propyne, can yield even better fits to the observed spectrum. \citet{Welbanks2025} also showcased how limited explorations of the vast model space and unaccounted degeneracies can lead to spurious claims of molecular detections.  

The structure of the paper is as follows. We present the Cassini data used in Section \ref{sec:titan_observation} and the models and retrieval framework in Section\,\ref{sec:titan_theory}. We refer the reader to \autoref{sec:likelihood} for the definitions and semantics adopted in this study, as they are central to the core message of the paper. The core findings of our retrievals for this application are given in Section\,\ref{sec:titan_results} with details in the Appendix. We discuss our findings in Section\,\ref{sec:discussion} and highlight major findings in Section\,\ref{sec:TitanConclusion}.

\begin{figure*}[ht!]
\vspace{+05mm}\includegraphics[width=1.05\linewidth, trim = 1cm 0.0cm 2cm 1cm, clip=true]{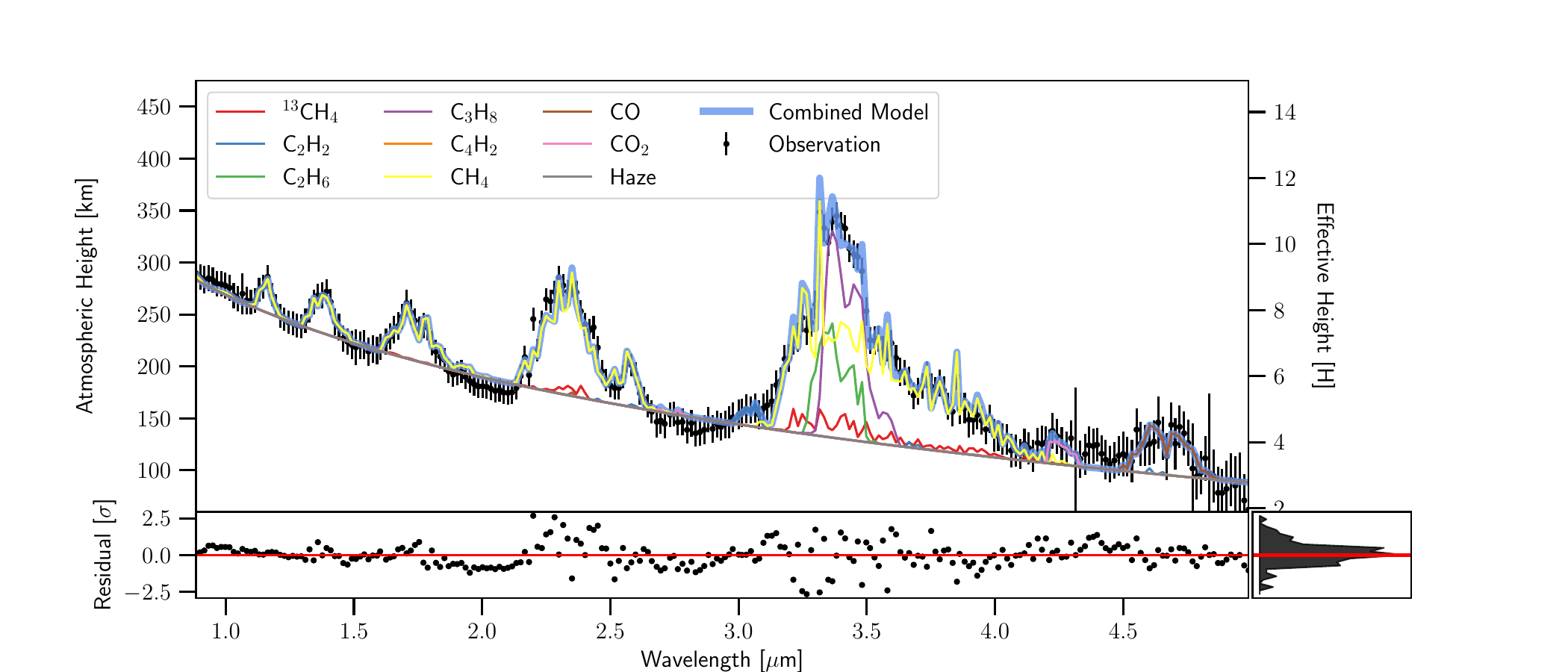}
\caption{A family portrait of hydrocarbons (HCs) in Titan's transmission spectrum obtained with Cassini/VIMS. Best fit model (thick blue line), corresponding to the maximum-likelihood solution, is shown for Titan’s spectrum obtained during visit T10 (see \autoref{tab:VisitTable}) from \citet{robinson2014}, together with the individual contributions of the absorbing molecules (see legend) that comprise of the molecular combination case 25 (see \autoref{tab:MolecularSpeciesOG}). The residuals, expressed in units of scaled observational uncertainty (see \autoref{eqn:likelihood}), are shown in the bottom-right panel along with their distribution. An Anderson–Darling test indicates a deviation from Gaussianity at the 5\% significance level. \label{fig:BestFitsTitan}}
\end{figure*}

\section{Observations}
\label{sec:titan_observation}
\citet{robinson2014} converted the VIMS observations of attitudinal transmission spanning from 0.88 $\mu$m to 5 $\mu$m into transmission spectra by numerically integrating over the line of sight. These observations show temporal variations between visits and are $\sim$10$\times$ more precise compared to \textit{JWST} observations with a similar wavelength coverage. Our retrievals are thus run on each of the four datasets (see \autoref{tab:VisitTable}).

\section{Models and Retrieval Framework}
\label{sec:titan_theory}
We run our in-house and publicly available transmission code \texttt{tierra} (\citet{niraula2022Nature} adapted from  \citet{dewit2013}). It uses a 1D formulation of the transmission spectroscopy commonly used in exoplanetary retrievals (see \citet{niraula2022Nature} for details) adding haze extinction on top of it. Previous work by \citet{robinson2014} has shown that the impact of the refraction for Titan is minimal, as the spectra probes altitudes greater than 100 km. Our retrievals are run with an assumption of an iso-mixture profile with free chemistry and with \textit{priors} of mixing ratio of between -12 to -1. It also assumes an isothermal profile owing to the fact that transmission spectroscopy primarily probes pressure levels beyond 100\,mbar corresponding to small temperature changes for objects far from their host star such as Titan. We ran additional tests with more complicated p-T profiles, but for the study, focus solely, on the result of the isothermal profile to underscore the key problem surrounding the retrievals of HCs. For the system parameters, we fix the mass to 0.0225  M$_\oplus$ and radius to 0.404 R$_\oplus$ adopted from NASA fact sheet\footnote{\href{https://nssdc.gsfc.nasa.gov/planetary/factsheet/saturniansatfact.html}{https://nssdc.gsfc.nasa.gov/planetary/factsheet/saturniansatfact.html}}.

\subsection{Molecular Species}
To assess the effect of \textit{a priori} molecule/model selection on the inferences reached, we perform our study using a series of molecular subsamples. Each molecule serves as an additional free parameter, thereby increasing the dimensionality of the retrieval problem. These molecules are selected taking into account the independent knowledge from Titan's atmosphere obtained from higher-resolution data (see \autoref{tab:MolecularSpecies}). The 25 sets of molecules are introduced in \autoref{tab:MolecularSpeciesOG}. As an example, in case 17 we fit for the abundances of \ce{CH4}, CO, \ce{CO2}, \ce{N2} (the ``Base Combination''),  C$_{2}$H$_{6}$,  C$_{3}$H$_{8}$, and C$_{2}$H$_{2}$. 

The rationale behind choosing the base combination of \ce{CH4}+\ce{CO}+\ce{CO2}+\ce{N2} is that these molecules exhibit largely non-overlapping spectral features, with \ce{N2} acting as the inert background gas. In total, we consider ten molecular species in our retrievals (see \autoref{tab:MolecularSpeciesOG}), a selection inspired by \citet{robinson2014}.

While, in principle, one could attempt all possible molecular combinations, the number of resulting models quickly becomes prohibitively large and computationally intractable. To address this, we adopt a hierarchical model-selection approach, whereby additional molecules are included in a structured manner to support our investigation cost-effectively. For example, the four-molecule base set is often insufficient to reproduce the complex spectral feature near 3.3~$\mu$m. This requires the inclusion of HCs such as ethane or propane. In the second family of retrievals, either ethane or propane is included, while for the third family of retrievals includes both ethane and propane to obtain better fitting models.

\begin{figure*}[ht!]
    \centering
    \vspace{03mm}\includegraphics[width=1.00\linewidth, trim = 0.1cm 0.1cm 0.1cm 0.2cm, clip=true]{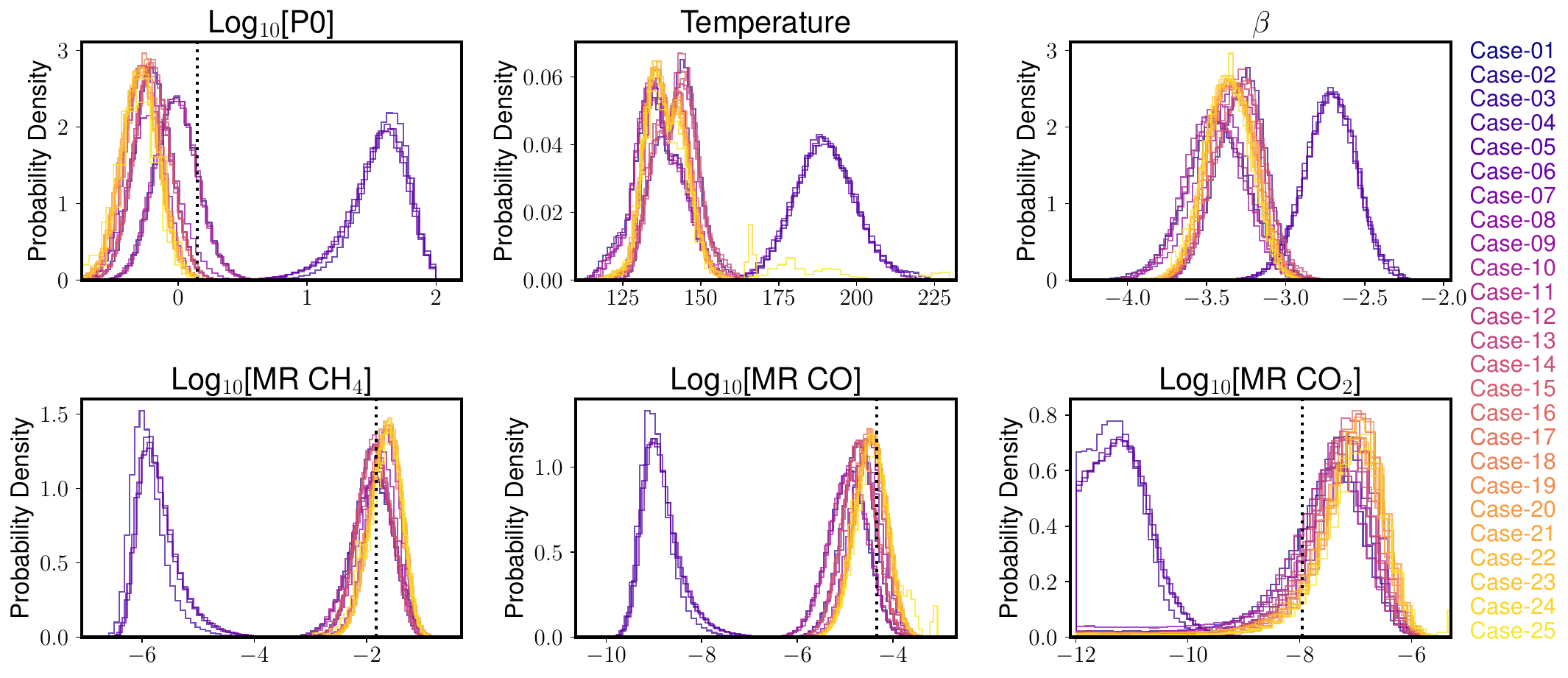}
    \caption{\label{fig:ppd_titan} Atmospheric inferences are sensitive to the molecules retrieved for. Posterior distribution of the base atmospheric parameters for an ensemble of 25 ``sets'', associated with different selections of molecules retrieved (see \autoref{tab:MolecularSpeciesOG}). The ``truths" reported in \autoref{tab:MolecularSpecies} are shown in the black dotted lines. Except for acetylene and carbon dioxide, the retrieved molecular values are generally consistent within 2$\sigma$ of previously reported values. Given the precision of the data, some extreme cases are highly unlikely. This is in particular the case of sets 3, 4, 5, and 6 that do not select a molecule with a sharp absorption feature at the center of the 3.3\,$\mu$m band forcing large compensations on the abundance of methane and other parameters while still leading exceedingly large structures ($\sim$50\,km vs $\sim$5\,km otherwise) in the residuals (see \autoref{fig:goodModelbadModel}). All other models lead to consistent fits and reveal a scatter of $\sim$0.5\,dex among the mean retrieved abundances for absorbers such as methane.}\vspace{01mm} 
    
\end{figure*}

\subsection{Opacity cross-sections}
We use HITRAN2020 \citep{hitran2020} line lists to calculate absorption cross sections for all molecules of interest except for propane, for which no line-by-line list exists in HITRAN. Thus, following \citep{Cours2020} we use the pseudo-line list for propane provided by Dr. Geoffrey Toon (JPL NASA)\footnote{\href{https://mark4sun.jpl.nasa.gov/pseudo.html}{https://mark4sun.jpl.nasa.gov/pseudo.html}}, which are based on the high-resolution laboratory measurements of \citet{harrison2010} that span the 3~$\mu$m region (i.e., 2560-3280 cm$^{-1}$). Cross sections for each molecule were calculated for a temperature range of 70 to 400 K with a stepsize of 14 K. Our pressure grid is the same as \citet{niraula2022Nature} value - spanning from 10$^{-5}$ to 100 bars.

\subsection{Haze opacity model}
Titan has a permanent, though temporally varying, haze layer that prominently impacts transmission spectroscopy \citep{lavvas2011, robinson2014}. Haze not only shrouds the surface, but it also mutes the transmission spectrum (see \autoref{fig:BestFitsTitan}). We use a similar formulation for the haze as in \citet{robinson2014}, assuming a wavelength-dependent power-law which is coupled with the atmosphere through the scale height ($H$) as $\tau(\lambda) = \tau_0 \cdot (\lambda)^{\beta} \cdot \exp\left({\frac{-z}{H}}\right)$,
where $\tau_0$ is the normalization factor, $\lambda$ is the wavelength in microns, $\beta$ is the power-law coefficient, $z$ is the height of the atmosphere from the reference radius, and $H$ is the atmospheric scale height.

\subsection{Retrieval Framework and Uncertainty Budget}
We use the retrieval framework introduced in \citet{niraula2022Nature} using the \textit{emcee} library. To account for sources of noises/uncertainties beyond the photon noise (e.g. instrumental systematics, imperfections in the opacity model, etc.), our framework maximizes the log of the likelihood \citep[e.g.,][]{Carter2009} (see \autoref{sec:likelihood}). We employed 100 walkers with a step size of 50,000. The evolution of the log-probability values was monitored to identify the burn-in phase, conservatively estimated at 25,000 step. The autocorrelation length was used to verify that the walkers were sufficiently mixed, indicating adequate convergence. The best-fit models presented in the following section correspond to the maximum likelihood solution.

\section{Results}
\label{sec:titan_results}
We present the results in this section that pertain to the detection-vs-retrieval challenges. More details on the ensemble of results are provided in the Appendix.

\subsection{Molecular Selection Shapes Retrievals}
\label{sec:detectable}

As shown in \autoref{fig:ppd_titan}, the choice of a subset of atmospheric compounds to be retrieved  (i.e., deemed ``detectable'') has an impact on more than the inference directly associated with them (such as their detection, their abundances, etc.). 
We find in particular that even the abundance of dominant species such as methane, can be affected by the selection of molecules because multiple species can contribute to the same spectral region, and retrievals may compensate by trading off one molecule’s abundance against another.

Given the precision of the data, some extreme cases are highly unlikely. This is in particular the result of cases 3, 4, 5, and 6 that do not select a molecule with a sharp absorption feature at the center of the 3.3\,$\mu$m band forcing large compensations on the abundance of methane and other parameters while still leading exceedingly large structures ($\sim$50\,km vs $\sim$5\,km otherwise) in the residuals (see \autoref{fig:goodModelbadModel} in the Appendix). However, the Anderson–Darling test for normality indicates that the residuals still deviate significantly from a Gaussian distribution at the 5\% significance level. Thus, any quantitative interpretation should be made with caution.

While the likelihood appears to favor the presence of ethane and propane, which also remains plausible as they are the next members in the HC series, without accurate cross-sections and the resulting non-Gaussian residuals means that robust conclusions cannot be drawn. This limitation is particularly critical in the 3.3 µm region, where numerous HCs exhibit overlapping features, making exhaustive model comparisons computationally prohibitive. Yet, among the retrievals that provide equivalent best fits (i.e. similar order of likelihood values), we find that the abundance of strong absorbers such as methane can be vary by $\sim$0.5\,dex (i.e., a factor of 3) and the temperature by up to 20\,K. Note that because the residuals are non-Gaussian we refrain for explicitly estimating and comparing the Bayes factor.

\section{Discussion}\label{sec:discussion}

\begin{figure}[ht!]
    \hspace{-10mm}
    \includegraphics[width=1.1\linewidth, trim = 1cm 1cm 1cm 1cm,clip=true]{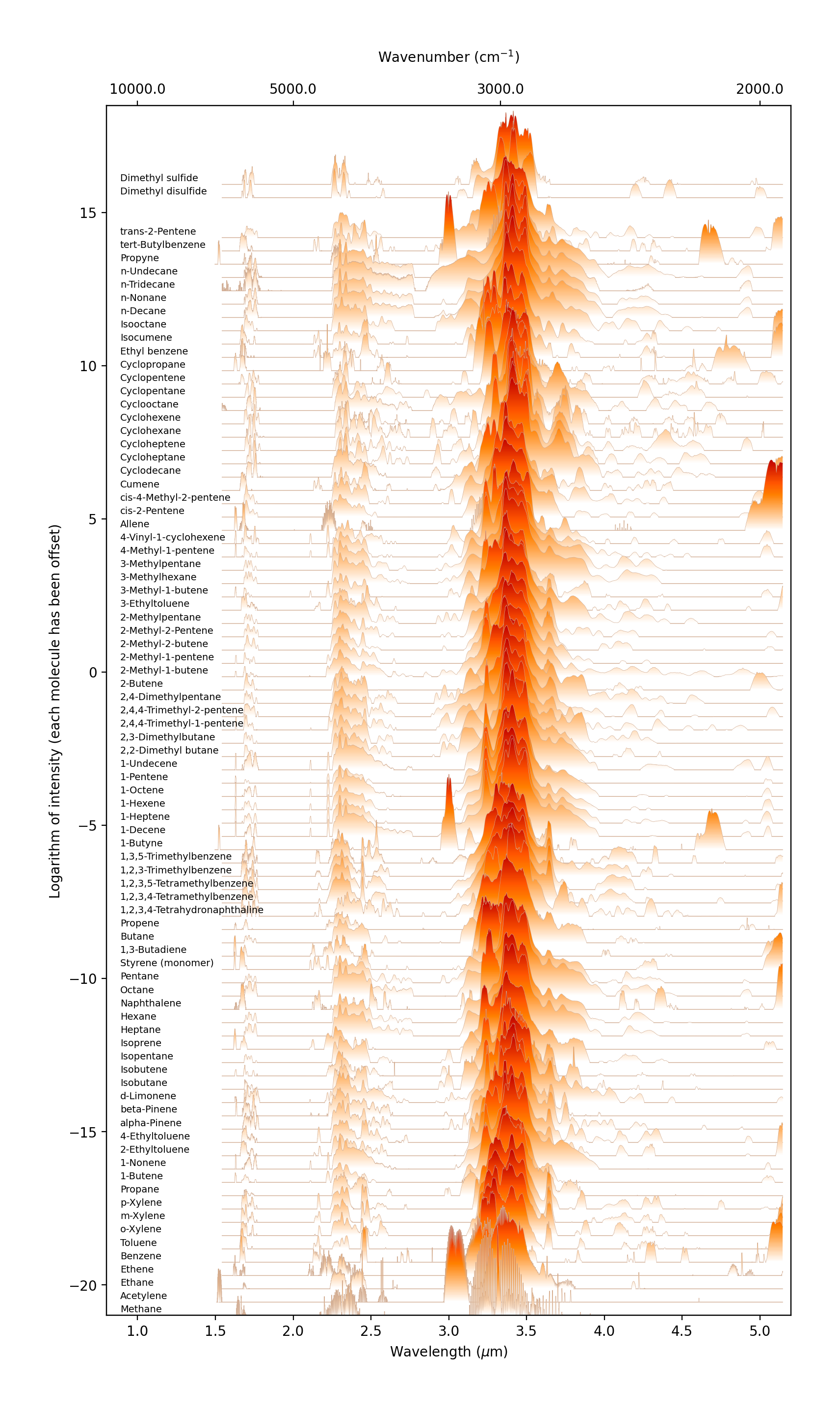}
    \caption{\label{fig:hydrocarbon} Spectra of different molecules may have many similarities, partially due to the presence of spectrally active shared functional groups \citep{sousaSilva2019, zhan2021assessment}. Overview of the absorption cross sections for HCs at 25$^{\circ}$C and 1.0~atm taken from PNNL \citep{PNNL_Sharpe_2004} and included in the cross sections part of the HITRAN2024 database \citep{XSC_HITRAN_2004}. The y-axis provides the logarithm of intensity, with each absorption cross section offset for display purposes. A minimum intensity of $1.0 \times 10^{-21}$~cm$^{2}$/molecule has been applied for each molecule.}
\end{figure}

\subsection{On the Similarities of Opacity Cross-Sections}
When obtained at high-resolution, high-SNR, and over a wide wavelength range, spectral information (e.g., in the form of cross-sections) can help identify compounds in exoplanet spectra. In the current observations, however,  the information content of the 3-3.5 $\mu$m band can be associated with thousands of molecules, including hundreds of HCs where this absorption feature is a consequence of a vibrational mode associated with the stretching symmetry of the C-H bond \citep{sousaSilva2019, zhan2021assessment}. \autoref{fig:hydrocarbon} presents infrared absorption cross sections for a set of HCs (and also dimethyl sulfide and dimethyl disulfide) taken from the Pacific Northwest National Laboratory (PNNL) database \citep{PNNL_Sharpe_2004} and now included in the ``Absorption cross sections" section (\url{www.hitran.org/xsc} of HITRAN2024 \citep{XSC_HITRAN_2004} (see further details in \autoref{fig:hydrocarbon2} in the Appendix). As shown, this selection of molecules mostly provides degenerate absorption features over the spectral range of \autoref{fig:BestFitsTitan}, with additional contributions from the combination and overtone bands near 2.2-2.6 $\mu$m and 1.5-1.7 $\mu$m. In addition, the 3.3 microns feature largely aligns with the signatures of DMS and DMDS due to the C-H bonds of the methyl group.  While the cross sections presented in \autoref{fig:hydrocarbon} and \autoref{fig:hydrocarbon2} are presented with a minimum intensity of $1.0 \times 10^{-21}$~cm$^{2}$/molecule, it should be noted that this has been applied for visual clarity, and these cross section measurements do include intensities below this limit.

When multiple molecules contribute to the same spectral band, a retrieval, to zeroth order, constrains the sum of their combined opacity. Consequently, the set of molecules included in the retrieval influences how spectral features are partitioned among species, thereby affecting the inferred abundances \citep[as also noted by][]{Cours2020}, and explaining the biases discussed in \autoref{sec:detectable}. A systematic evaluation of all species consistent with the observed features is essential, but the vast parameter space involved makes robust HC retrievals challenging (e.g., \citealt{Welbanks2025}, Fig.~1). Moreover, meaningful model comparisons require accurate and comprehensive molecular cross-sections, many of which are unavailable or incomplete, forcing us to rely on intermediate-precision datasets (e.g., RASCALL \cite{sousaSilva2019}) , a practice which not only highlights model vulnerabilities but also further introduce uncertainties to the retrieval process.

Complementary information, including insights from atmospheric chemistry, can narrow plausible atmospheric compositions. Yet, it is crucial to assess model precision, especially understanding the impact of imprecise opacities. Sensitivity analyses akin to \citet{niraula2022Nature} for opacities  and \citet{broussard2025impact} for photochemistry, where perturbations are applied at the level of estimated model precision, can help quantify the reliability and interpretive limits of retrievals.

\begin{figure*}[ht!]
    \centering
    \includegraphics[width=0.95\linewidth, trim = 0.1cm 0.1cm 1cm 0.25cm,clip=true]{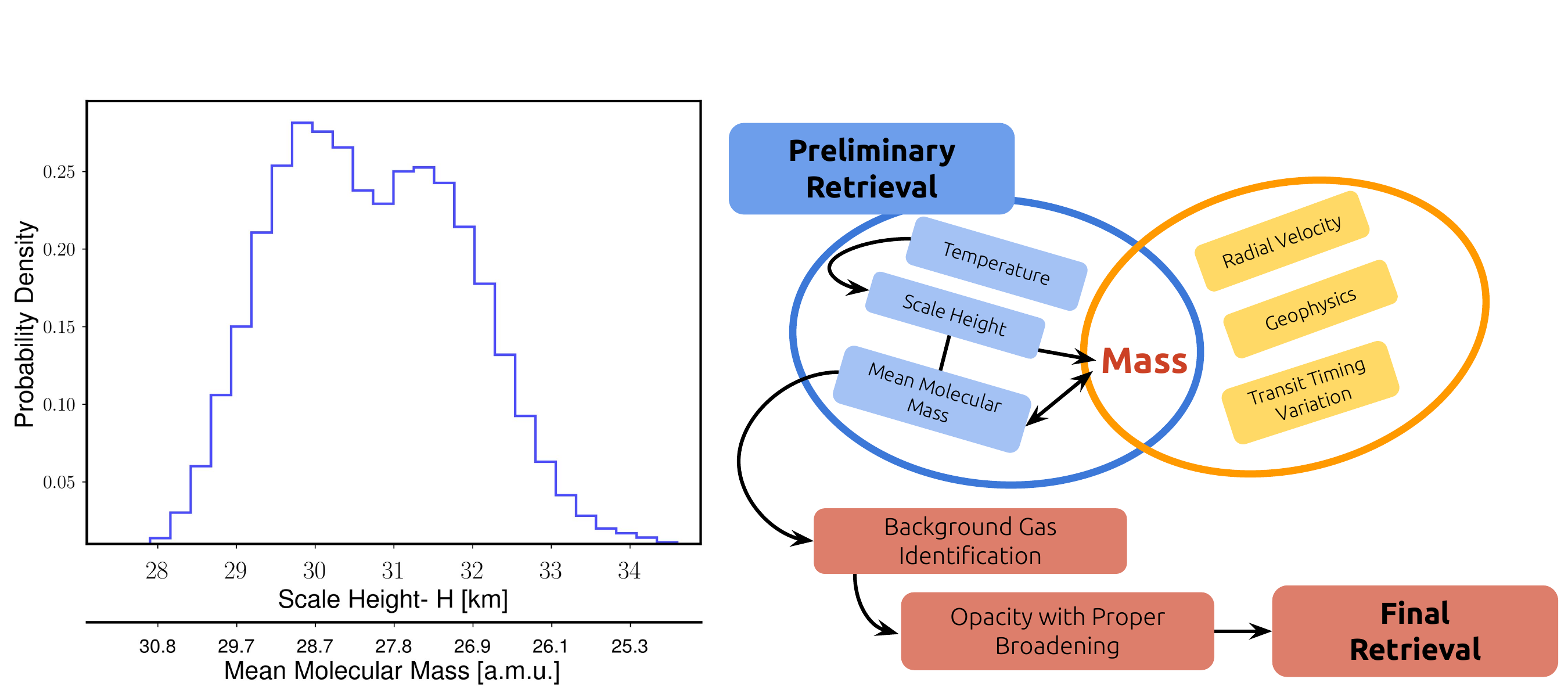}
\caption{\label{fig:TitanFrameworkTerrestrialPlanet} Background gas identification via constraints on the scale height from transmission spectroscopy. \textbf{Left:} PPD of  Titan's atmospheric scale height yielded by its transmission spectrum (consistent with the stratospheric value). \textbf{Right} Framework to yield robust abundances for terrestrial planets using iterative retrievals through identification of background gas.  
}
\end{figure*}

\subsection{A Special Case: Background Gas Identification}

Nitrogen makes up the bulk of Titan's atmosphere and yet does not exhibit prominent absorption features in its spectrum. As a result, the major constituents of Titan's atmosphere were debated for more than a decade, until a mixture of nitrogen and methane was suggested based on arguments of Jean's escape and spectroscopic observations \citep{hunten1973}. Such conclusions were arrived at through models taking advantage of spatially resolved temperature measurements. Therefore, when it comes to distant terrestrial worlds, identifying the underlying background gas will likely be difficult due to degeneracies introduced by clouds \citep{benneke2013, welbanks2019} though not impossible---e.g., through weak collision-based absorption (CIA) features \citep{kaltenegger2020, welbanks2019}). 

Fortunately, retrievals from transmission spectra can yield a strong constraint on the atmospheric pressure scale height that, combined with independent constraints on the object's mass and radius, can yield the mean molecular mass as an indicator of the main atmospheric compound \citep{benneke2012}. \citet{dewit2013} had previously demonstrated that the atmospheric scale height, temperature, and composition can be constrained independently from transmission spectroscopy--which can yield constraints on the planetary mass. Amongst the synthetic scenarios explored, \citet{dewit2013} included cases where the dominant atmospheric constituent was a weak absorber (such as $\rm{H_2}$ and $\rm{N_2}$), yet either their scattering slope or their CIA was detectable, as very hazy atmospheres were not included, resulting in a tight constraint on the mean molecular mass.

Our retrievals yield $H = 30.6\pm1.6$\,km consistent with Titan's stratospheric scale height (\autoref{fig:TitanFrameworkTerrestrialPlanet}). Combined with Titan's gravity, this translates into constraints on a mean atmospheric molecular weight of 27.8 $\pm$ 1.8 a.m.u ($\mu = kT/gH$, with $k$ Boltzmann's constant, $T$ the local temperature, $g$ the local gravity, and $H$ the scale height). As none of the strong absorbers identified are dominant compounds (abundances $\ll$10\%), the most likely weak-absorber gas matching the derived molecular weight is Nitrogen. 

For exoplanet atmospheric studies, we thus suggest following an iterative retrieval process (see \autoref{fig:TitanFrameworkTerrestrialPlanet}). This involves first inferring the background gas through scale height (with complementary insights form radial-velocity measurements and/or transit-timing variations were possible), and using the inferred background gas for subsequently generating high(er)-fidelity opacity models \citep[e.g., with adequate broadening parameters,][]{niraula2022Nature, Wiesenfeld2025}.

We note that for a large regime of planets, distinguishing between a light (e.g., H$_2$) and a heavy (e.g., N$_2$) weakly-absorbing dominant atmospheric component could be possible without independent knowledge on the planetary mass. This is enabled by the fact that their mean molecular weights are one order of magnitude apart. Therefore, for a given scale height constrained via transmission spectroscopy, they each relate to a planetary mass also separated by an order of magnitude -- and one of them is likely nonphysical (marked as ``Geophysics'' in \autoref{fig:TitanFrameworkTerrestrialPlanet}). In the present case, without any $a~priori$ knowledge about Titan's mass, from its atmospheric scale height and temperature alone, one can reject a low-$\mu$ atmosphere as it would imply a mass 10 times its actual mass, meaning a density $\sim$3.5 times Earth's, which is non-physical.

However, both high mean molecular atmospheric signals and high-altitude gray clouds (including photochemical hazes) suppress the transmission spectral features, and in practice, as we struggle with SNR of transmission spectra for such planets and their atmospheres, this has proven to be a challenge. These challenges are also further exacerbated by the degeneracy of mass with the molecular abundances \citep{benneke2013, line2016, batalha2019, welbanks2019}.  In presence of such clouds, retrievals are also further impacted by cloud-parametrization used in the models \citet{barstow2020}. Nevertheless, there have been an increasing number of low molecular weight atmospheres that turned to transmission spectroscopy to constrain their mass \citep[see][]{deWit2025Nature}.

\section{Conclusion}
\label{sec:TitanConclusion}

The high-fidelity data from and independent knowledge of Titan can be used to contextualize characterization frameworks in the new era of exoplanetary data. We found in particular that:

\begin{enumerate}

    \item The choice of molecules retrieved for in exoplanetary studies can lead to significant biases (0.5-0.75\,dex) on the abundance of constituents associated with overlapping absorption features.
    \item Titan's transmission spectrum being anchored (from an information standpoint) by the 3.3\,$\mu$m feature is a challenge when combined with the fact that hundreds of HCs (amongst other molecules) has a strong absorption feature in this region.
    \item The last point, combined with the limited or scarce opacity data for ``heavy'' HCs (heavier than propane), is a bottleneck for performing retrievals in Titan's atmosphere, and will likely be for future exo-Titans. 
    \item We recommend a careful assessment of the ensemble of molecules matching detected spectral features, while leveraging complementary insights (e.g., from atmospheric chemistry) to reduce the range of possibilities and, finally, performing a detailed sensitivity analysis to assess the dependencies of atmospheric inferences on the final ensemble of possible molecule/model selections.
    \item While several molecules (methane, ethane, propane) can be consistently retrieved across four different visits, this ``retrieval" preference cannot be interpreted as bonafide ``detection." 
    \item Finally, we show that identifying the dominant atmospheric constituents without relying on absorption features (e.g., for weak absorbers such as nitrogen or hydrogen) can be done through constraints on the scale height (and thus possible mean molecular weights) to support adequate opacity models (e.g., adequate broadening parameters) and a broader understanding of the planetary environment studied. 
\end{enumerate}

\vspace{5mm}
\facilities{Cassini}

\textit{Softwares}:
\texttt{tierra}\footnote{\url{https://github.com/disruptiveplanets/tierra}}, \texttt{emcee}

\acknowledgments
The authors thank Keeyoon Sung for an updated opacity model for propane, and Sara Seager \& Sai Ravela for their valuable inputs. P.N. and J.d.W. acknowledge support from the European Research Council (ERC) Synergy Grant under the European Union’s Horizon 2020 research and innovation program (grant No. 101118581 — project REVEAL). The authors acknowledge the MIT SuperCloud and Lincoln Laboratory Supercomputing Center for providing (HPC, database, consultation) resources that have contributed to the research results reported within this paper. 

\clearpage

\bibliography{Bibliography}

\appendix

\section{Additional Results \& Discussions}

We present here additional findings that are beyond the primary scope of this Letter and, yet, still of relevance to the community. This includes the visit-to-visit consistency of our findings, the effect and modeling of hazes, and parameter correlations.

\subsection{Likelihood and Model Selection}
\label{sec:likelihood}

The likelihood function employed in our retrievals is given by:
\begin{align}
    \mathcal{L} &= -\frac{1}{2}\sum\left[\frac{(\text{y}-\mathcal{M}(\theta))^2}{\sigma_T^2} + \text{ln} \, 2 \pi \sigma_T^2\right] \\
    \sigma^2_T &= \sigma_{obs}^2 + f^2\sigma_{obs}^2,
    \label{eqn:likelihood}
\end{align}

where $\mathcal{M}(\theta)$ is the model spectrum given parameters $\theta$, $\text{y}$ is the observed spectrum, and $\sigma_{obs}$ denotes the observational uncertainties, scaled here by an error inflation factor $f$, analogous to a jitter term used in radial velocity analyses, which accounts for underestimated or unmodeled noise (for full treatment of noise using co-variance matrix see \citet{rotman2025}).  Within the Bayesian framework, this likelihood is combined with the \textit{a priori} (or prior) distribution, $p(\theta)$, which is often informed by chemistry. And combined with likelihood, this can then be used for estimating the posterior distribution $(p(\theta \mid \text{y}))$ for a given set of observations (y):
\[
p(\theta \mid \text{y}) \propto \mathcal{L}(\text{y} \mid \theta) \, p(\theta).
\]

If $\mathcal{M}_N$ and $\mathcal{M}_{N+1}$ represent models with $N$ and $N+1$ molecules respectively, the detection of an additional molecule is commonly assessed through comparison of their Bayesian evidence (see \citet{benneke2013}). However, this approach assumes that the model errors are small and the residuals are Gaussian. Uncertainties in molecular cross-sections often lead to a violation of this assumption. In particular, the cross-sections of HCs vary significantly in both accuracy and completeness. For instance, the cross-sections for methane are comparatively well determined, those for ethane are less so, and those for propane are even more limited. For more specialized molecules like DMS/DMDS, only a handful of measurements are present, that too at standard pressure and temperature.

Ignoring errors from the models can lead to the apparent ``retrieval'' of a molecule (e.g. \citet{Madhusudhan2025}) i.e. a model being preferred simply because the evidence seems to favor it.  However, claiming this ``retrieval" as a genuine ``detection" overlooks the vast number of possible scenarios and does not necessarily represent the optimal explanation of the data. Hence, what may appear as a statistically supported retrieval claim could instead stem from incomplete model physics,  or cross-sectional degeneracies, among other limitations. \citet{Madhusudhan2025} ``detection'' of DMS/DMDS in K2-18 b suffers from this critical caveat. While they compare model with and without DMS/DMDS, their molecular selection only included a small subset of hydrocarbons and their canonical baseline model did not explore a wider range of molecular combinations. When a wider range of hydrocarbons are included, the evidence for DMS/DMDS  gets significantly weaker as discussed under Section ``Interpretation According to Conventional Practices" and their Figure 5 in \citet{Welbanks2025}.

\subsection{Findings Across 4 Visits}
Our retrievals specifically lead to detection of multiple HCs (CH$_{\rm 4}$, C$_{\rm 2}$H$_{\rm 6}$, C$_{\rm 3}$H$_{\rm 8}$, and C$_{\rm 2}$H$_{\rm 2}$), and CO (\autoref{fig:BestFitsTitan}). 
As shown in \autoref{fig:ppd_titan_visits}, our retrieved parameters are consistent within 2$\sigma$ with the values reported in the literature (\autoref{tab:MolecularSpecies}). 
Across four different epochs (\autoref{tab:VisitTable}),  the retrieved parameters yield relatively consistent insights.  

The primary differences across visits relates to the haze parameters and the abundances of secondary components such as CO$_2$. As shown in \autoref{fig:ppd_haze}, the contribution of Titan's haze in visit T53 is significantly larger than in visit T10, leading to weaker spectral features (such as CO$_2$'s 4.3$\,\mu$m feature) to be muted. With the data at hand, we cannot attribute the changes in the abundances of C$_{\rm 3}$H$_{\rm 8}$ to variability in the atmosphere across visits or different pressure levels being probed due to the effect of hazes (the changes for C$_{\rm 2}$H$_{\rm 6}$ and C$_{\rm 2}$H$_{\rm 2}$ are not significant).

As we refer to a possible dependence of our findings on the pressure level probed, we note that acetylene (C$_{\rm 2}$H$_{\rm 2}$) is found with an abundance 3$\sigma$ larger than expected values from \citet{coustenis2016}. We suspect this arises from the photochemical production of acetylene in the upper atmosphere, the region most probed in transmission, resulting in overestimated abundances when uniformly extrapolated to deeper, unprobed layer. However, this could also arise from the current model limitations including but not limited to isothermal profile.

Similarly, the abundance of methane is known to reach as high as 5.65\% $\pm$ 0.18\% near Titan's surface \citep{niemann2010} while stratospheric oscillates around 1.48$\pm$0.09\% \citep{horst2017} -- consistent with our retrievals. The values reported in \autoref{tab:MolecularSpecies} are adopted from Table 1 of \citet{sylvestre2018}, which, similar to our retrievals,  normalizes the total volume mixing ratio to unity by digesting observations/retrievals from multiple sources. While, these values lack temporal, spatial as well as altitudinal specificity, and chemical species such as acetylene and ethane can vary by an order of magnitude or more \citet{coustenis2016}. 

\subsection{Haze Profile}
Our current treatment of the haze profile with a single power law appears adequate for our retrievals. However, if we are to extend the transmission spectroscopy to the far ultraviolet (see \citet{tribbett2021}), the current single power law fails to capture the opacity trend. Haze has a prominent impact on Titan's transmission spectra. It shows a clear spatiotemporal variation, which can be caused by seasonally varying atmospheric circulation patterns \citep{rannou2010}. The strong molecular absorption features (such as from methane) are identifiable on the top of the hazes; however, hazes often dominate the absorption continuum, masking any weaker spectral features that could potentially have been observed in between the methane bands (e.g. CO).

\subsection{Correlation Among Parameters}
\autoref{fig:TitanCornerPlot} is a corner plot for Titan for visit T53 for the isothermal model for molecular set 7. A number of parameters show a strong correlation, including the haze normalization factor ($\tau_0$), the power-law coefficient ($\beta$), and retrieved temperature (T$_{\rm 0}$). However, the retrieved molecules are uncorrelated with the hazes, showing that the strong absorbers in these atmospheres can still be characterized by their molecular features despite the presence of atmospheric hazes. The correlation among the abundances of retrieved parameters is driven by their dependency on the scale height \citep{dewit2013, niraula2022Nature}. The strongest correlation is seen between methane and propane, likely due to their overlapping spectral features. This is expected to be a concern, not just between methane and propane, but across HC detections, many of which share spectral features associated with the C-H bond vibration and its overtones \citep{zhan2021assessment}. High spectral resolutions, combined with complete spectral information for all relevant HCs, are likely to be required to unambiguously distinguish their individual contributions to Titan's atmosphere, or any future observations of HC-rich atmospheres.

\newpage
\section{Additional Tables \& Figures}

\begin{figure*}[ht!]
    \centering
    \includegraphics[width=0.95\textwidth, trim = 0cm 0cm 0cm 0cm, clip=true]{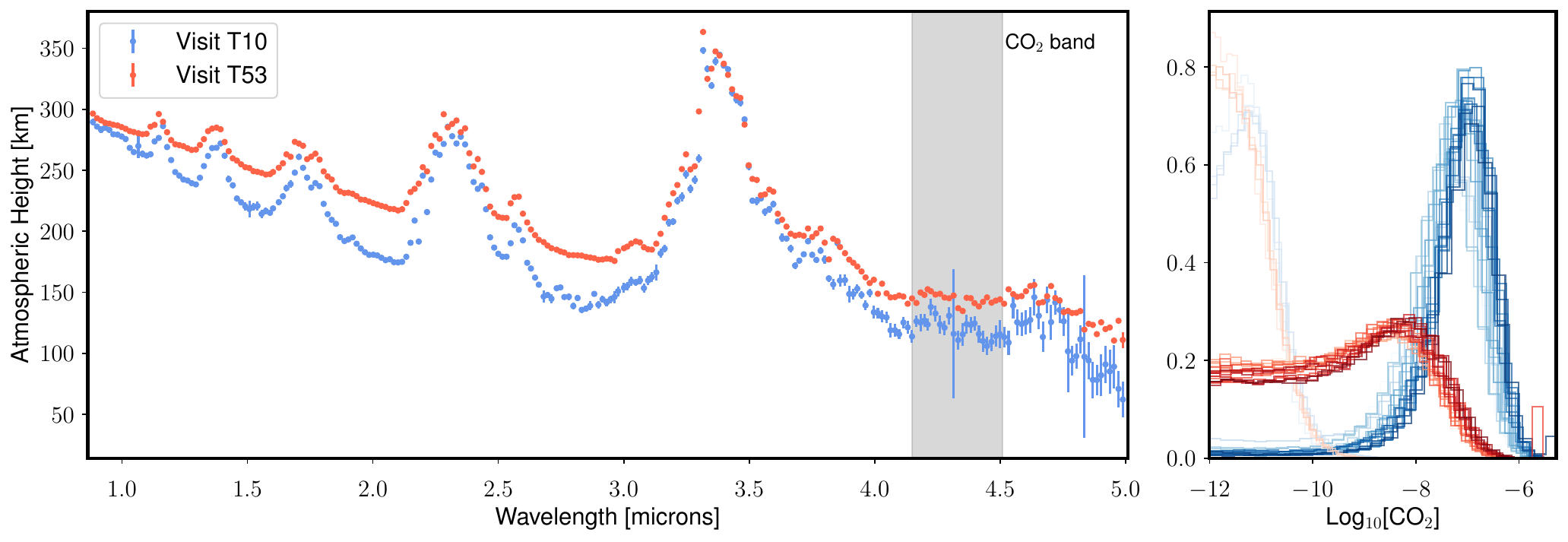}\caption{\label{fig:ppd_haze} Weak absorption features muted during hazy epochs. \textbf{Left:} Titan's Transmission spectrum corresponding to two different visits (T10 and T53, see \autoref{tab:VisitTable}). T53 is affected by hazes to such an extent that weak absorption features such as CO$_2$'s at 4.3\,$\mu$m are mostly muted. \textbf{Right:} Retrieved CO$_{2}$ mixing ratios for T10 and T53 showing that in the latter case the detection of CO$_2$ is significantly less significant due to hazes muting its 4.3\,$\mu$m band.}
\end{figure*}

\begin{figure*}[ht!]
    \centering
    \includegraphics[width=0.95\textwidth, trim = 0cm 0.25cm 0cm 0.25cm,clip=true]{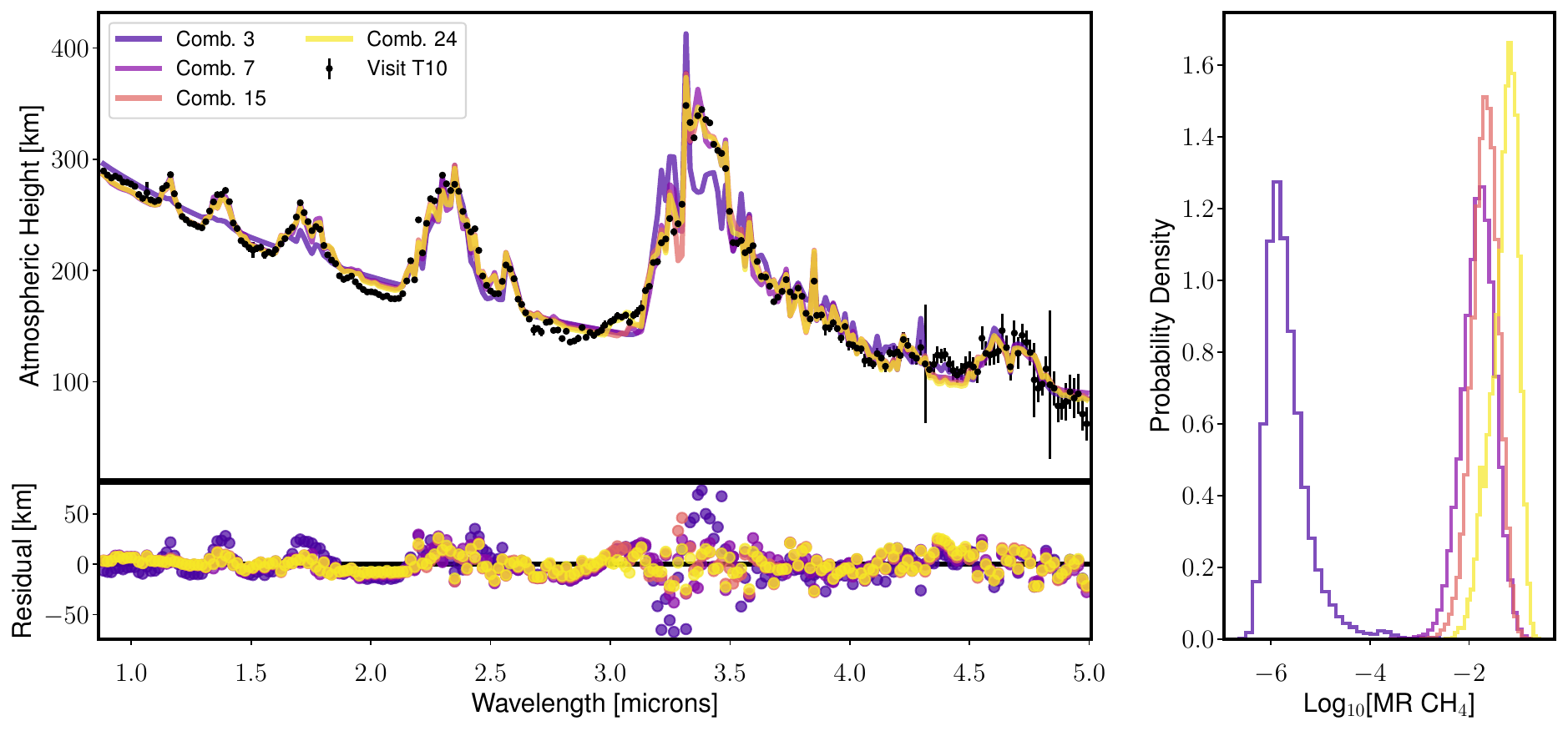}
    \caption{\label{fig:goodModelbadModel} Not all choices of molecule sets yield equally good fits. Fits of four different combinations of models with the corresponding residuals (left) and posterior distribution of methane's mixing ratio (right). Colors indicate the sets of molecules used for each retrieval (see \autoref{tab:MolecularSpeciesOG}). Sets of molecules without ethane and/or propane lead to substantially worse fits. All other sets lead comparable fits, yet a range of methane mixing ratio spreading over 0.5-0.75\,dex is associated with such pre-retrieval selections.}
\end{figure*}

\begin{figure*}[ht!]
    \centering
    \includegraphics[width=0.98\textwidth, trim = 0.1cm 0.1cm 0.1cm 0.1cm, clip=true]{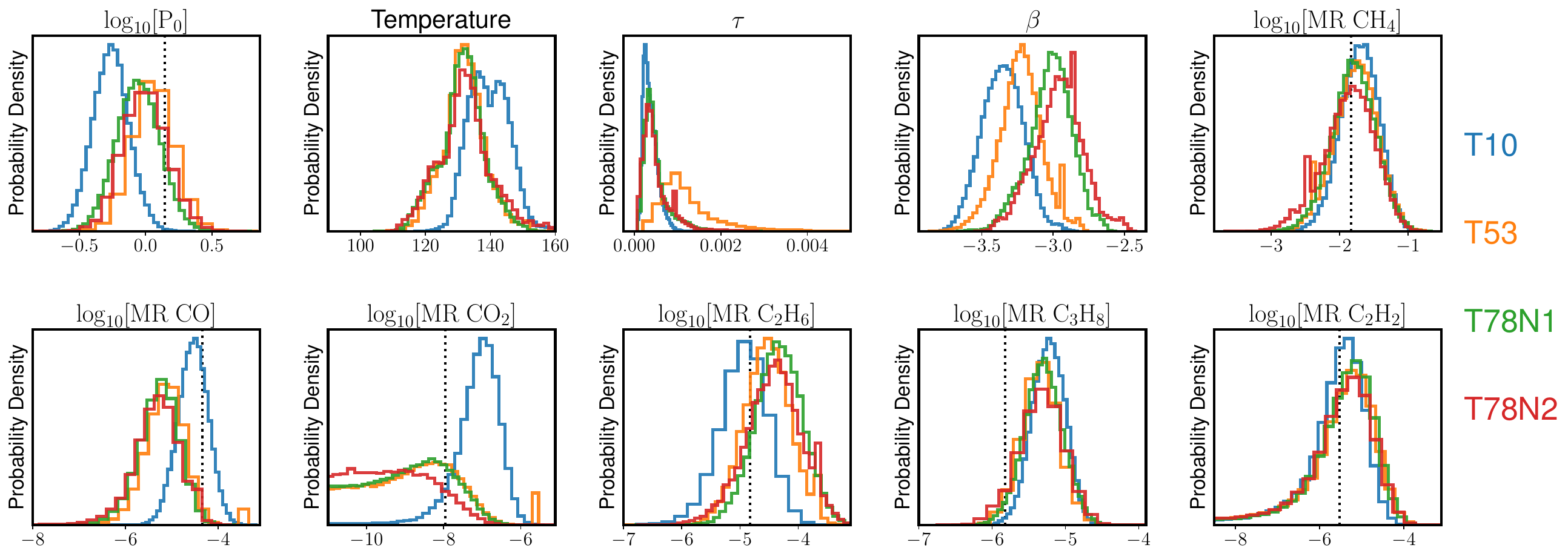}
    \caption{\label{fig:ppd_titan_visits} Posterior distribution of different molecules observed for Titan in our retrievals for molecular set 17. The ``truths" reported in \autoref{tab:MolecularSpecies} are shown in the black lines. Except for acetylene and carbon dioxide, the retrieved molecular values are generally consistent within 2$\sigma$ of previously reported values.}
\end{figure*}

\begin{figure*}[t!]
    \includegraphics[width=1.0\textwidth, trim = 0.cm 0cm 0cm 0.25cm,clip=true]{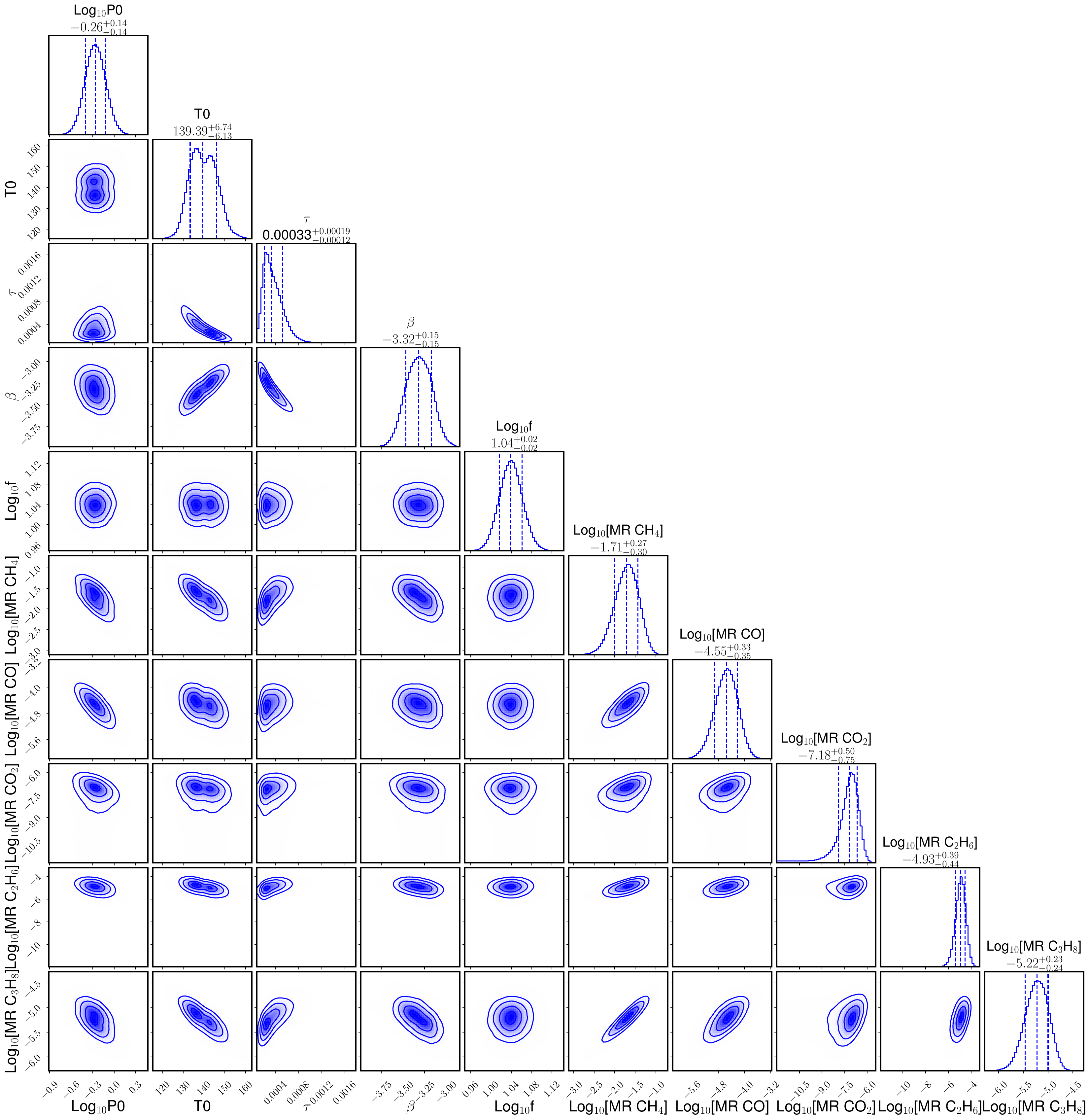}
    \caption{\label{fig:TitanCornerPlot} Corner plot showing correlation among different parameters for visit T10 for combination set of 7. A number of parameters, including retrieved abundance of methane and retrieved abundance of propane, show a strong correlation.}
\end{figure*}

\begin{figure}[ht!]
    \hspace{-10mm}
    \includegraphics[width=1.1\linewidth, trim = 1cm 1cm 1cm 1cm,clip=true]{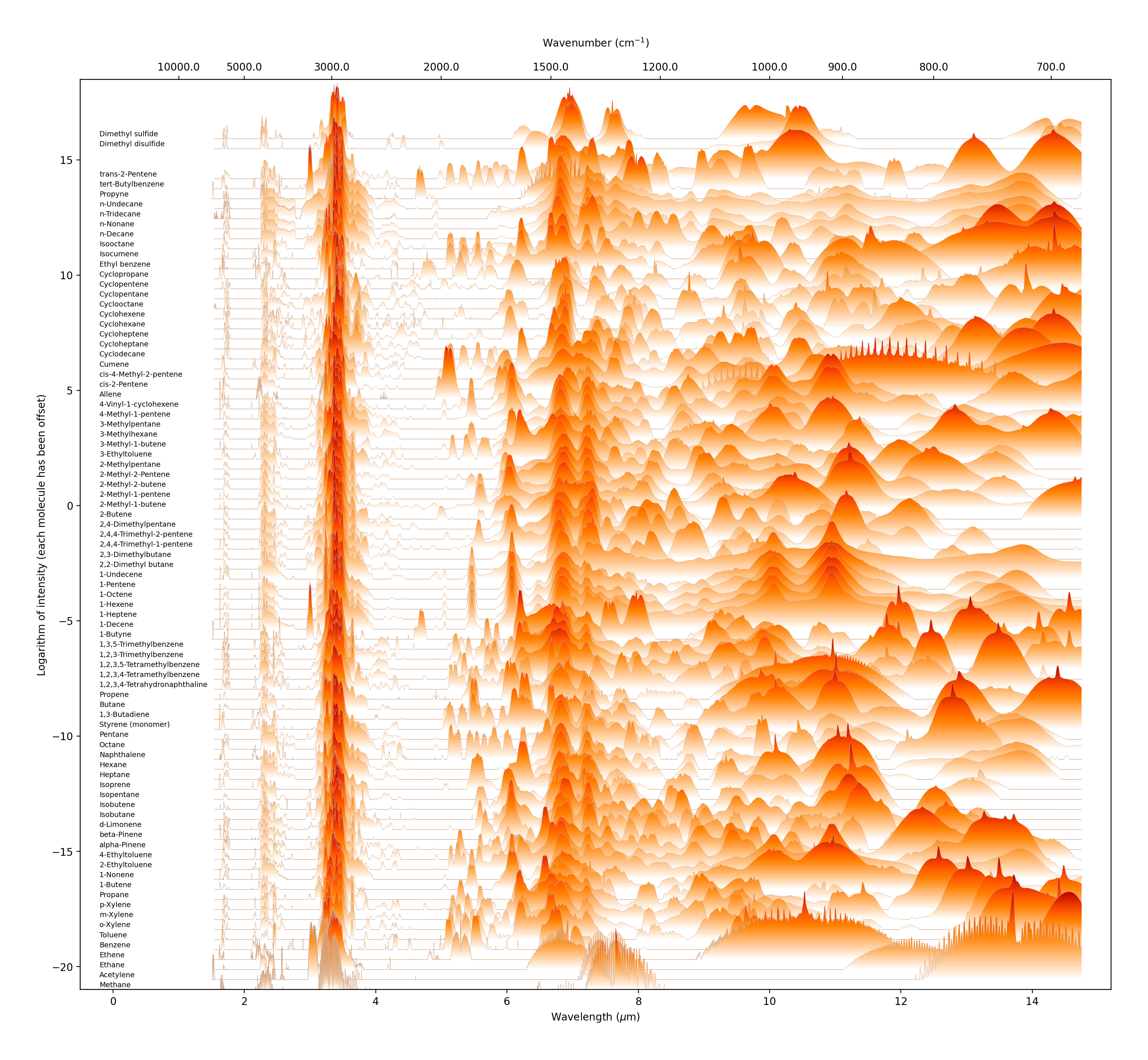}
    \caption{\label{fig:hydrocarbon2}Overview of the absorption cross sections for HCs, DMS and DMDS, taken from PNNL \citep{PNNL_Sharpe_2004} and included in HITRAN2024 database \citep{XSC_HITRAN_2004}. Each measurement is at 25$^{\circ}$C, 1.0~atm, and has been convolved to a 1.0~cm$^{-1}$ resolution. The y-axis provides the logarithm of intensity and each molecule has been offset for display purposes. A minimum intensity of $1.0 \times 10^{-21}$~cm$^{2}$/molecule has been applied for each molecule for the purpose of visual clarity. The full spectral range of the underlying PNNL data is presented, showing the degeneracy below 5 $\mu$m (also shown in \autoref{fig:hydrocarbon}). This degeneracy is partially lifted above 5 $\mu$m, in the mid- to far-IR, due to the differences in the low frequency vibrational modes of the C-C and C-H bonds.}
\end{figure}

Here we provide additional relevant details regarding our retrievals. \autoref{tab:VisitTable} lists all data sources used in this analysis, while \autoref{tab:MolecularSpeciesResults} summarizes the molecular abundances adopted from \citet{sylvestre2018}, which are used to benchmark our retrieved values. \autoref{fig:ppd_haze} presents retrieval results for all model configurations across two independent visits, T10 and T53. \autoref{fig:goodModelbadModel} illustrates four different cases of retrieved methane abundances for various model combinations, highlighting the wide range of inferred values. \autoref{fig:ppd_titan_visits} shows the posterior distributions of retrieved atmospheric molecular abundances from four different visits, and \autoref{fig:TitanCornerPlot} presents the corresponding corner plot for visit T10.  \autoref{fig:hydrocarbon2} provides an overview of the absorption cross-sections of hydrocarbons, illustrating the strong degeneracies present, particularly at wavelengths shorter than 5 $\mathrm{\mu}$m.

\begin{table}[!h]
    \caption{Cassini/VIMS data adapted from \citet{robinson2014} used in this work.  \label{tab:VisitTable}}
    \centering
    \begin{tabular}{r|c|c|c}
    \hline
    \textbf{Cassini Flyby} &\textbf{Season}& \textbf{Latitude}& \textbf{Date}\\
    \hline
     T10 & N Winter & 70$^\circ$ South & Jan. 2006\\
     T53 & Equinox & 1$^\circ$ North & Apr. 2009 \\
     T78N1 & N Spring & 27$^\circ$  &  North Sep. 2011\\
     T78N2 & N Spring & 40$^\circ$  & North Sep. 2011 \\
    \hline
    \end{tabular}
\end{table}

\vspace{-2cm}
\input{tables/volumeMixingRatio}
\input{tables/retrievalCombination}

\end{document}

%% file: tables/volumeMixingRatio.tex
\begin{table*}
    \caption{Reported volume mixing ratios for molecules present in the atmosphere of Titan  \label{tab:MolecularSpeciesResults}}
    \centering
    \begin{tabular}{r|c|c|c}
    \hline
    \textbf{Molecular Species}& Volume Mixing Ratio
    & Source & Cross-section$^{\S}$ \\
    \hline
    Nitrogen [$\mathrm{N_2}$]$^\dagger$ & 0.9839 & Normalization & H20 \\
    Methane [$\mathrm{CH_4}$]$^\dagger$ & 0.0148 & \citet{niemann2010}& H20 \\
    Methane [$\mathrm{^{13}CH_4}$]$^\ddagger$ & 1.71$\times$10$^{-4}$ & \citet{nixon2012}& H20 \\
    Carbon Monoxide [$\mathrm{CO}$]$^\dagger$ & 4.6$\times$10$^{-5}$ & \citet{maltagliati2015}& H20 \\
    Ethane [$\mathrm{C_2H_6}$]$^\dagger$ & 1.5$\times$10$^{-5}$ & \citet{coustenis2016}& H20 \\
    Propane [$\mathrm{C_3H_8}$]$^\ddagger$ & 1.5$\times$10$^{-6}$ & \citet{coustenis2016}& NJPL \\
    Acetylene [$\mathrm{C_2H_2}$]$^\dagger$& 3.0$\times$10$^{-6}$ &  \citet{coustenis2016}& H20 \\
    Ethylene [$\mathrm{C_2H_4}$]$^\ddagger$ & 1$\times$10$^{-7}$ & \citet{coustenis2016}& H20 \\
    Carbon dioxide [$\mathrm{CO_2}$]$^\ddagger$ & 1.1$\times$10$^{-8}$ & \citet{coustenis2016}& H20 \\
    Diacetylene [$\mathrm{C_4H_2}$]$^\ddagger$ & 2.0$\times$10$^{-9}$ & \citet{teanby2009}& H20 \\
    Benzene [$\mathrm{C_6H_6}$]$^\ddagger$ & 4.0$\times$10$^{-10}$ & \citet{coustenis2016}& --- \\
    \hline
    \end{tabular}
 \vspace{1ex}
{\par{\raggedleft }}
{\raggedright $^\dagger$Values as reported in \citet{sylvestre2018}\par}
{\raggedright $^{\S}$ Where reference codes refer to: H20 -- HITRAN2020 \citep{hitran2020}, NJPL -- NASA JPL pseudo-line list based on the measurements of \citet{harrison2010}\par}

\label{tab:MolecularSpecies}
\end{table*}

%% file: tables/retrievalCombination.tex
\begin{table*}
    \caption{Sets of molecules used for the retrievals\label{tab:MolecularSpeciesOG}. Base Combination \ce{CH4}+\ce{CO}+\ce{CO2}+\ce{N2} is used, onto which additional combination of molecules are added. In the second category of molecular selection, either ethane or methane is used and in the third category both of them are used.}
    \centering
    \begin{tabular}{r|l||c|l||r|l}
    \hline
    \textbf{Set}& \textbf{Molecules} & \textbf{Set}& \textbf{Molecules} &\textbf{Set}& \textbf{Molecules+ C$_{2}$H$_{6}$ + C$_{3}$H$_{8}$}   \\
      \hline  
1. & C$_{2}$H$_{6}$ & 9. & C$_{2}$H$_{6}$ + C$_{2}$H$_{2}$ & 17. &  C$_{2}$H$_{2}$  \\
2. & C$_{3}$H$_{8}$ & 10. & C$_{2}$H$_{6}$ + C$_{2}$H$_{4}$ &  18. &  C$_{2}$H$_{4}$\\
3. & $^{13}$CH$_{4}$ &11. & C$_{2}$H$_{6}$ + C$_{4}$H$_{2}$  & 19. &  C$_{4}$H$_{2}$ \\
4. & C$_{2}$H$_{2}$ & 12. & C$_{3}$H$_{8}$ + $^{13}$CH$_{4}$ &  20. &  $^{13}$CH$_{4}$ + C$_{2}$H$_{2}$\\
5. & C$_{2}$H$_{4}$ & 13. & C$_{3}$H$_{8}$ + C$_{2}$H$_{2}$ &  21. & $^{13}$CH$_{4}$ + C$_{2}$H$_{4}$ \\
6. & C$_{4}$H$_{2}$ & 14. & C$_{3}$H$_{8}$ + C$_{2}$H$_{4}$ &  22. & $^{13}$CH$_{4}$ + C$_{4}$H$_{2}$ \\
7. & C$_{2}$H$_{6}$ + C$_{3}$H$_{8}$ & 15. & C$_{3}$H$_{8}$ + C$_{4}$H$_{2}$ & 23. & $^{13}$CH$_{4}$ + C$_{2}$H$_{2}$ + C$_{2}$H$_{4}$\\
8. & C$_{2}$H$_{6}$ + $^{13}$CH$_{4}$ & 16. & C$_{2}$H$_{6}$ + C$_{3}$H$_{8}$ + $^{13}$CH$_{4}$ & 24. & $^{13}$CH$_{4}$ + C$_{2}$H$_{2}$ + C$_{4}$H$_{2}$ \\
& & & & 25. & $^{13}$CH$_{4}$ + C$_{2}$H$_{2}$ + C$_{4}$H$_{2}$ + C$_{2}$H$_{4}$  \\
\hline
\end{tabular}
\vspace{1ex}
\end{table*}